\documentclass[a4paper,12pt]{article}
\usepackage[T1]{fontenc} 
\usepackage{amsmath} 
\usepackage{amssymb} 
\usepackage{bm} 

\DeclareMathOperator{\e}{\mathnormal{e}}
\DeclareMathOperator{\diag}{diag}

\begin{document}

\begin{center}
\textbf{\large Cosmology constrains gravitational four-fermion interaction}
\end{center}

\begin{center}
I.\,B. Khriplovich$^{a, b}$ \footnote{khriplovich@inp.nsk.su} and A.\,S. Rudenko$^{a, b}$ \footnote{a.s.rudenko@inp.nsk.su}
\end{center}
\begin{center}
$^a$ Budker Institute of Nuclear Physics, 630090, Novosibirsk, Russia \\
$^b$ Novosibirsk State University, 630090, Novosibirsk, Russia
\end{center}

\vspace{1mm}

\begin{abstract}
If torsion exists, it generates gravitational four-fermion interaction (GFFI). This interaction gets dominating on the Planck scale. If one confines to the regular, axial-axial part of this interaction, the results do not comply with the Friedmann-Lema\^{i}tre-Robertson-Walker (FLRW) cosmology for the spatial flat or closed Universe. In principle, the anomalous, vector-vector interaction could restore the agreement.
\end{abstract}

\vspace{3mm}

\textbf{1.} The observation that, in the presence of (non-propagating) torsion, the interaction of fermions with gravity results in the four-fermion interaction of axial currents, goes back at least to \cite{ki, ro}.

We start our discussion of GFFI with the analysis of its most general form. This interaction looks as follows: 
\begin{equation} \label{ff} 
S_{ff} = \frac{3 \pi G}{2} \frac{\gamma^2}{\gamma^2 + 1} \int d^4 x \, \sqrt{- g} \, \eta_{I J} \left[ A^I A^J + \frac{\alpha}{\gamma} \left( V^I A^J + A^I V^J \right) - \alpha^2 \, V^I V^J \right]; 
\end{equation} 
here and below $G$ is the Newton gravitational constant; $g$ is the determinant of the metric tensor; $A^I$ and $V^I$ are the total axial and vector neutral currents, respectively:
\begin{equation} \label{A} 
A^I = \sum_a A_a^I = \sum_a \bar \psi_a \, \gamma^5 \, \gamma^I \, \psi_a \, , \qquad V^I = \sum_a V_a^I = \sum_a \bar \psi _a \, \gamma^I \, \psi_a \, , 
\end{equation} 
the sums over $a$ in \eqref{A} extend over all sorts of elementary fermions with spin 1/2. As to the so-called Barbero-Immirzi parameter $\gamma$, its numerical value $\gamma = 0.274$ \cite{kk} is the solution of the "secular" equation 
\begin{equation}
\sum_\text{$j = 1/2$}^\infty \left( 2 j + 1 \right) e^{- 2 \pi \gamma \sqrt{j (j + 1)}} = 1. 
\end{equation} 
$\alpha$ is a free parameter of the problem, its numerical value is unknown.

The $AA$ contribution to expression \eqref{ff} corresponds (up to a factor) to the action derived long ago in  \cite{ki, ro}. Then, this contribution was obtained in the limit $\gamma \to \infty$ in \cite{ke} (when comparing the corresponding result from \cite{ke} with \eqref{ff}, one should note that our convention $\eta_{IJ} = \diag \left( 1, - 1, - 1, - 1 \right)$ differs in sign from that used in \cite{ke}). The present form of the $AA$ interaction, given in \eqref{ff}, was derived in \cite{pe}. The $VA$ and $VV$ terms in \eqref{ff} were derived in \cite{fr}.

Simple dimensional arguments demonstrate that interaction \eqref{ff}, being proportional to the Newton constant $G$ and to the particle number density squared, gets essential and dominates over the common interactions only at very high densities, i.\,e. on the Planck scale and below it.

The list of papers where the GFFI is discussed in connection with cosmology, is too lengthy for this short note. Therefore, we refer here only to the review \cite{be}, with a quite extensive list of references (in all of them the discussion is confined to the analysis of the axial-axial interaction).

\vspace{3mm}

\textbf{2.} We will be interested in the energy-momentum tensor (EMT) $T_{\mu \nu}$ generated by action \eqref{ff}. Therein, the expression in square brackets has no explicit dependence at all either on the metric tensor, or on its derivatives. The metric tensor enters action $S_{ff}$ via $\sqrt{- g}$ only, so that the corresponding EMT is given by relation 
\begin{equation} 
\frac{1}{2} \int d^4 x \, \sqrt{- g} \; T_{\mu \nu} = \frac{\delta}{\delta g^{\mu \nu}} \, S_{ff} \, . 
\end{equation} 
Thus, with identity 
\begin{equation}
\frac{1}{\sqrt{- g}} \, \frac{\delta \sqrt{- g}}{\delta g^{\mu \nu}} = - \frac{1}{2} \, g_{\mu \nu} \, , 
\end{equation} 
we arrive at the following expression for the EMT \cite{khr0, khr}: 
\begin{equation} \label{EMT} 
T_{\mu \nu} = - \frac{3 \pi G}{2} \frac{\gamma^2}{\gamma^2 + 1} \, g_{\mu \nu} \, \eta_{I J} \left[ A^I A^J + \frac{\alpha}{\gamma} \left( V^I A^J + A^I V^J \right) - \alpha^2 \, V^I V^J \right]. 
\end{equation} 
The nonvanishing components of this expression, written in the locally inertial frame, are energy density $T_{00} = \rho$ and pressure $T_{11} = T_{22} = T_{33} = p$ (for the correspondence between $\rho$, $p$ and EMT components see \cite{ll}, \S\,35).

Thus, the equation of state (EOS) is here\footnote{Note, that in Refs. \cite{khr0, khr} the $AA$ contribution to $\rho$ and $p$ is 4 times smaller. AR thanks S.\,K. Maity for the query that led to a recalculation of $AA$ and $VV$ contributions.}
\begin{equation} \label{EOS} 
\rho = - \, p = - \, \frac{\pi}{12} \, \frac{\gamma^2}{\gamma^2 + 1} \; G n^2 \left[ \left( 3 - 11 \, \zeta \right) - \alpha^2 \left( 15 - 7 \, \zeta \right) \right]. 
\end{equation} 
In this expression, $n$ is the total density of fermions and antifermions, and $\zeta = \langle \bm \sigma_a \bm \sigma_b \rangle$ is the average value of the product of corresponding $\bm \sigma$-matrices, presumably universal for any $a$ and $b$. Since the number of sorts of fermions and antifermions is large, one can neglect here for numerical reasons the contributions of exchange and annihilation contributions, as well as the fact that if $\bm \sigma_a$ and $\bm \sigma_b$ refer to the same particle, $\langle \bm \sigma_a \bm \sigma_b \rangle = 3$.

It is only natural that after the performed averaging over all momenta orientations, the $P$-odd contributions of $VA$ to $\rho$ and $p$ vanish.

The parameter $\zeta$\,, just by its physical meaning, in principle can vary in the interval from 0 (which corresponds to the complete thermal incoherence or to the antiferromagnetic ordering) to 1 (which corresponds to the complete ferromagnetic ordering). However, it looks quite reasonable to assume that, at the discussed extremal conditions of high densities and high temperatures, this correlation function $\zeta$ is negligibly small. Then, EOS \eqref{EOS} simplifies to
\begin{equation} \label{EOS1} 
\rho = - \, p = - \, \frac{\pi}{4} \, \frac{\gamma^2}{\gamma^2 + 1} \; G n^2 \, \left( 1 - 5 \mspace{1 mu} \alpha^2 \right) \, . 
\end{equation}

\vspace{3mm}

\textbf{3.} Let us assume now that, even on the scale where EOS reduces to \eqref{EOS}, the Universe is homogeneous and isotropic, and thus is described by the well-known Friedmann-Lema\^{i}tre-Robertson-Walker (FLRW) metric 
\begin{equation} \label{fr} 
ds^2 = dt^2 - a(t)^2 \left[ dr^2 + f(r) \left( d\theta^2 + \sin^2 \theta\, d\phi^2 \right) \right]; 
\end{equation} 
here $f(r)$ depends on the topology of the Universe as a whole:
\[
f(r) = r^2, \quad \sin^2 r, \quad \sinh^2 r
\]
for the spatial flat, closed, and open Universe, respectively.

The Einstein equations for the FLRW metric \eqref{fr} reduce now to 
\begin{align} 
\left( \frac{\dot a}{a} \right)^2 + \frac{k}{a^2} & = \frac{8 \pi G \rho}{3} \, , \label{a1} \\
\frac{\ddot a}{a} = - \frac{4 \pi G}{3} \left( \rho + 3 p \right) & = \frac{8 \pi G \rho}{3} \, . \label{a2} 
\end{align} 
Parameter $k$ in eq. \eqref{a1} equals 0, 1, and $- 1$ for the spatial flat, closed, and open Universe, respectively.

In fact, observational data strongly favor the idea that our Universe is spatial flat, i.e. that $k = 0$ in formula \eqref{a1}. Then, the energy density $\rho$ should be positive. So much the more, $\rho$ should be positive if the Universe is closed, i.\,e. if $k = 1$. However, if one confines to the canonical, $AA$ structure of the GFFI, i.\,e. omits anomalous $VV$ term in formula \eqref{EOS1}, $\rho$ is negative which is certainly unsatisfactory.

Equations \eqref{a1} and \eqref{a2} are supplemented by the covariant continuity equation, which can be written as follows: 
\begin{equation} \label{e} 
\dot \rho + 3 H \left( \rho + p \right) = 0, \quad H = \frac{\dot a}{a} \, . 
\end{equation} 
For the energy-momentum tensor \eqref{EOS1}, dominating below the Planck scale, and resulting in $\rho = - p$, this equation \eqref{e} reduces to 
\begin{equation} \label{e1} 
\dot \rho = 0 \quad \Rightarrow \quad \rho = \text{const}. 
\end{equation} 
In this way, for $\rho > 0$, we arrive with \eqref{a2} and \eqref{e1} at the following expansion law: 
\begin{equation} \label{e2} 
a \sim \e^{H_0 t}, \quad \mathrm{where } \ H_0 = \sqrt{\frac{8 \pi G \rho}{3}} = \text{const} 
\end{equation} 
(as usual, the second, exponentially small, solution of eq. \eqref{a2} is neglected here). Thus, GFFI results here in the inflation starting below the Planck scale. Obviously, in this case as well, the reasonable physical result takes place only for $\rho > 0$.

To summarize, GFFI can result in reasonable physical conclusions, if any, only in the presence of the anomalous $VV$ interaction, i.\,e. for $\alpha^2 > 1/5$, see \eqref{EOS1}.

\vspace{3mm}

\textbf{4.} In conclusion, let us come back to EMT \eqref{EMT}, which can be rewritten as 
\begin{equation} 
T^\mu_\nu = \delta^\mu_\nu \, \rho \, . 
\end{equation} 
As long as this contribution to the total EMT dominates below the Planck scale, it should be conserved by itself. In this way we arrive at 
\begin{equation}
\partial_\nu \mspace{1 mu} \rho = 0 \, , \qquad \text{i.\,e.} \quad \dot \rho = 0 \, , \quad \nabla \rho = 0 \, . 
\end{equation} 
Thus, here the energy density and pressure, $\rho = - \, p$\,, are constant both in time and coordinates. In other words, EMT \eqref{EMT} by itself has no dynamics at all.

\subsection*{Acknowledgements}

The investigation was supported in part by the Foundation for Basic Research through Grant No. 11-02-00792-a, by the Ministry of Education and Science of the Russian Federation, and by the Grant of the Government of Russian Federation, No. 11.G34.31.0047.

\end{document}